# ROBUST WAVE FUNCTION OPTIMIZATION PROCEDURES IN QUANTUM MONTE CARLO METHODS

Dario Bressanini[a] and Gabriele Morosi[b]
*Dipartimento di Scienze Chimiche, Fisiche e Matematiche. Universita' dell'Insubria Sede di Como, via Lucini 3, 22100 Como (Italy)*

Massimo Mella[c]
*Dipartimento di Chimica Fisica ed Elettrochimica, Universita' di Milano, via Golgi 19, 20133 Milano (Italy)*

The energy variance optimization algorithm over a fixed ensemble of configurations in variational Monte Carlo is formally identical to a problem of fitting data: we reexamine it from a statistical maximum-likelihood point of view. We detect the origin of the problem of convergence that is often encountered in practice and propose an alternative procedure for optimization of trial wave functions in quantum Monte Carlo. We successfully test this proposal by optimizing a trial wave function for the Helium trimer.

## INTRODUCTION

The problem of the optimization of a trial function with many nonlinear parameters for a quantum system is still an open issue. This is especially true in the field of quantum Monte Carlo simulations, where usually one uses trial wave functions for which the analytical evaluation of the energy is impossible. For this reason, the variational Monte Carlo (VMC) method is used to numerically evaluate the energy and other properties of the trial wave functions, and to optimize them. While in standard quantum mechanical calculations it is common to optimize a trial wave function by minimizing its variational energy, in VMC simulations it is much more common to minimize the variance of the local energy $\sigma^2(H) = \langle H^2 \rangle - \langle H \rangle^2$, rather than the energy itself. The reason is that the minimization of the energy by a Monte Carlo method is much more troublesome from a numerical point of view than the minimization of its variance, which is however far from being problem-free, as discussed below.

The minimization of the variance, as an alternative to the minimization of the energy, has been first proposed, to the best of our knowledge, by Weinstein[1] in the context of lower bound calculations of eigenvalues of the Schrödinger equation, and by Bartlett *et al.*[2], who suggested the use of the variance of the local energy as a criterion of goodness for approximate wave functions. Frost[3-7] and Conroy[8-10] implemented the minimization of the variance within a numerical scheme to optimize trial wave functions for small atoms and molecules. Variance minimization was introduced in VMC by Coldwell[11] and Umrigar and coworkers[12]. The main reason this method is rarely applied nowadays in standard computational methods is that its use involves the expectation value of the square of the Hamiltonian, a very difficult quantity to compute analytically. A second problem is that for some (albeit, very poor) trial wave functions this quantity might diverge. A third problem is that sometimes, again for very poor wave functions, a local minimum cannot be found (for example, it is easy to check that using a single Gaussian function to describe the ground state of the Hydrogen atom, a minimum of $\sigma^2(H)$ does not exist). The main point, i.e. the difficulty in its computation, is not an issue in VMC, while the other two problems are usually not met in practical calculations. However the issue of the quality of the optimized wave function remains to be settled. For an infinitely flexible trial wave function, both energy and variance minimization procedures recover the exact ground state, in practice however, using an incomplete basis, the resulting wave functions are different. Whether or not the variance optimized wave function is an overall better wave function than the energy optimized one has been discussed in the QMC literature[13-15]. Initially the variance optimized wave function was thought to be at least as good as the energy optimized one, if not superior for certain properties. At present there are suggestions that this is not so[13-16], especially if the trial function is not flexible enough. These conclusions seem to confirm the results that James and Coolidge[17] and later Goodisman[18] obtained a long time ago in the field of ab initio methods. For this reason, several papers have recently focused on the problem of minimizing the energy using VMC[14,19]. However, the modern energy optimization algorithms can approach only few electron systems and in any case the methods do not appear to be as fast as the standard variance minimization algorithm. For these reasons the variance minimization method is still generally used in VMC and it easy to predict that its use will last. In spite of this large diffusion, there is a problem the quantum Monte Carlo practitioner is sometimes faced while trying to optimize a trial wave function using a fixed ensemble of configurations. Varying the parameters in the trial wave function to minimize the variance of the local energy, the optimization algorithm reaches a minimum and produces the best set of parameters for that fixed ensemble of configurations. However, when the new trial wave function is used in a VMC simulation, often it appears even worse than the previous function.

---

[a] Electronic mail: Dario.Bressanini@uninsubria.it
[b] Electronic mail: Gabriele.Morosi@uninsubria.it
[c] Electronic mail: Massimo.Mella@unimi.it

We call *false convergence* this problem we are addressing in this paper. We need to understand its origin in order to make the algorithm more reliable and able to generate better wave functions. The origin of the problem can be traced back to the presence in the fixed ensemble of some configurations that have values of the local energy very different from the average. We call these walkers "bad" as their presence in the fixed ensemble spoils the minimization process. A practical, and sometimes effective, solution is to eliminate these "bad" walkers from the fixed ensemble in some *ad-hoc* way. For example retaining only those walkers whose local energy is within a fixed chosen window, or discarding all the walkers whose local energy is off by $N\sigma$ from the average value, for some value of N, as suggested by Kent et al.[20]. These solutions are far from being satisfactory from the theoretical point of view, since these empirical criteria are somehow arbitrary. A better solution would be to design a method that automatically deals with these bad walkers in a non-arbitrary way.

In order to understand why things sometimes go wrong and how to correctly deal with bad walkers, we first need to understand what we are *really* doing when we are minimizing the variance of the local energy. We will show that the origin of the problem of false convergence might not reside in the bad walkers, but rather in the optimization procedure.

Suppose we have N data points $(\mathbf{R}_i, E_L(\mathbf{R}_i))$ where $\mathbf{R}_i$ represents the walker and $E_L(\mathbf{R}_i)$ its local energy associated to the trial wave function $\Psi(\mathbf{R};\mathbf{a})$ that depends on the adjustable parameters vector $\mathbf{a}$. To minimize the variance of the local energy with respect to some reference energy, close to the exact energy, means to find the minimum

$$\min_{\mathbf{a}} \sum_i^N (E_L(\mathbf{R}_i, \mathbf{a}) - E_R)^2 \qquad (1)$$

where we have explicitly shown the dependence of the local energy on the parameter vector $\mathbf{a}$.

We now derive Eq. 1 in the standard way, and show that an alternative derivation, focused on the discrete algorithm used to search the minimum, involving a sum over discrete points and not a continuous integration, can shed light on the problem of false convergence.

## *The integral point of view*

The usual way to derive Eq. 1 is to start from the variance minimum principle[3,8,12,21]

$$\sigma^2(H) = \langle H^2 \rangle - \langle H \rangle^2 \geq 0 \qquad (2)$$

For an eigenstate, the variance of the Hamiltonian is at a local minimum, and equal to zero.
Optimizing a wave function using this principle means solving the problem

$$\min_{\mathbf{a}} \sigma^2(H(\mathbf{a})) \qquad (3)$$

Explicitly writing the integrals

$$\min_{\mathbf{a}} \sigma^2(H(\mathbf{a})) = \min_{\mathbf{a}} \frac{\int \Psi(\mathbf{R};\mathbf{a})^2 (E_L(\mathbf{R};\mathbf{a}) - \langle H \rangle)^2 d\mathbf{R}}{\int \Psi(\mathbf{R};\mathbf{a})^2 d\mathbf{R}} \qquad (4)$$

In practice, since the integrals cannot be evaluated analytically, they are estimated using a finite (small) number of integration points, generated by VMC. Suppose the VMC process generated N points $\mathbf{R}_i$ distributed according to $\Psi(\mathbf{R};\mathbf{a})^2$, the discrete approximation of the variance is

$$\sigma^2(H(\mathbf{a})) \cong \frac{1}{N} \sum_i^N (E_L(\mathbf{R}_i;\mathbf{a}) - \langle H \rangle)^2 \qquad (5)$$

Instead of minimizing $\sigma^2(H(\mathbf{a}))$, sometimes it is preferable to minimize a related quantity, namely the second moment of the local energy with respect to a fixed (or reference) energy $E_R$,

$$\mu^2_{E_R}(H(\mathbf{a})) = \frac{\int \Psi(\mathbf{R};\mathbf{a})^2 (E_L(\mathbf{R};\mathbf{a}) - E_R)^2 d\mathbf{R}}{\int \Psi(\mathbf{R};\mathbf{a})^2 d\mathbf{R}} \cong \frac{1}{N} \sum_i^N (E_L(\mathbf{R}_i;\mathbf{a}) - E_R)^2 \qquad (6)$$

The constant $E_R$ should be close to the energy of the state being sought, although the optimization does not depend strongly on its value. Minimizing this quantity (which many authors call $\sigma^2(H(\mathbf{a}))$ without making any distinction) is almost equivalent to minimizing the variance. A little algebra shows that

$$\mu^2_{E_R}(H) = \sigma^2(H) + (\langle H \rangle - E_R)^2 \qquad (7)$$

from which Eq. 1 is recovered, using the minimum principle of Eq. (2).

This is, in short, what we might call *the integral point of view*. Adopting this view, one uses Eq. 5 or Eq. 6. Taking the integral point of view, it is very easy to invent different functionals upon which to base the minimization process. For examples

$$\frac{\int (E_L(\mathbf{R}_i;\mathbf{a}) - E_R)^2 \Psi(\mathbf{R})^4 d\mathbf{R}}{\left(\int \Psi(\mathbf{R})^2 d\mathbf{R}\right)^2} \qquad (8)$$

and

$$\frac{\int |E_L(\mathbf{R}_i;\mathbf{a}) - E_R| \Psi(\mathbf{R})^2 d\mathbf{R}}{\int \Psi(\mathbf{R})^2 d\mathbf{R}} \qquad (9)$$

and several others have been suggested and discussed by Alexander *et al.*[22], in the context of the biased-selection Monte Carlo method. In general, any functional of the general form

$$\frac{\int P(\mathbf{R}) f(E_L(\mathbf{R}_i;\mathbf{a}) - E_R) d\mathbf{R}}{\int P(\mathbf{R}) d\mathbf{R}} \qquad (10)$$

can be proposed, as long as $P$ is a probability distribution and $f$ is a non-negative function such that $f(0)=0$.

Although this derivation of the optimization process is clean and simple, it does not shed any light on the problems of false convergence that, as mentioned before, one often encounters in practice. In the derivation, the only role played by the discretization process is to approximate the integrals by finite sums, while the focus is really on the integral functional. However it is really the discretization process that is the source of the problem. So, if we want to clarify this issue, we must start from a point of view that includes the discrete nature of the algorithm from the very beginning.

Furthermore, why should one use the variance functional (Eq. 5 and 6) rather the functionals of Eq. 8 and 9, or one of the many possible others? Is there any *a priori* reason to prefer one to the others?

## *The fitting point of view*

If the discretization process would be only a way to approximate the integrals involved in the different functionals, we should be very surprised by the usually good quality of the results obtained with the generally rather small number of points used (usually from some hundreds to some thousands). As correctly pointed out by Umrigar and coworkers[12], there is really no need to invoke any kind of integral approximation to justify the algorithm. The key observation is that the correct way to look at the optimization algorithm is as a fitting process, and the minimization of Eq. 1 should be considered in its own right, with no connection to integrals. This crucial point was already pointed out by Frost[4] who observed that, if $\Psi$ is the exact wave function and $E_R$ is the exact energy, the sum in Eq. 1 is zero for *any* distribution of points, whether or not they are chosen to yield good approximations to integrals. Since the discrete nature of the algorithm is intrinsic in the *fitting point of view*, from this perspective we can hope to clarify the problem of false convergence. A related analysis, starting from a different point of view, of the effects of approximating the required integrals by finite sums has been recently carried out by Kent and coworkers[20].

If we do not start by invoking the variance of the Hamiltonian, we must ask ourselves why should we minimize the square of the deviation of the local energy rather than its absolute deviation, or its fourth power or other formulas. Which criterion should drive us to select one among them? The answer, well known to statisticians, is the *maximum likelihood estimator*. We can view the above formulas as fitting criteria of N measurements, affected by errors. Each algorithm generates the vector **a** that best reproduces these values. There are vectors **a** that are very "unlikely" and others that are more "likely" since they give an average energy closer to $E_R$. We now *assume* that these values are affected by normally distributed independent errors with a fixed standard deviation σ. The *likelihood* of a single observation is

$$P_i \propto e^{-\frac{1}{2}(E_L(\mathbf{R}_i;\mathbf{a}) - E_R)^2 / \sigma^2} \qquad (11)$$

The total probability, assuming the same value of σ for each point, is

$$P_{tot} \propto \prod_i^N e^{-\frac{1}{2}(E_L(\mathbf{R}_i;\mathbf{a}) - E_R)^2 / \sigma^2} \qquad (12)$$

The usual (and often implicit) choice at this point is to select the set of parameters **a** such that the total probability (or likelihood) is maximized. This is equivalent to maximizing the logarithm of Eq. (12), leading to Eq. 1.

So, the familiar least squares fitting in Eq. 1 is a maximum likelihood estimation of the fitted parameters if the underlying distribution is assumed to be a Gaussian. Implicit is the assumption that points far from the average value are very unlikely, since the tails of a Gaussian distribution decay very fast. These points, when they are present, influence the least-squares procedure more than they ought. In the statistical jargon these points are called *outliers*. If they occur they might bias the fitting procedure to produce meaningless values of the parameters. The problem lies in the fact that the probability of these points in the assumed Gaussian model is so small that the maximum likelihood estimator tries to distort the whole model to take them into account.

Now that we have identified the problem, we need to understand why we encounter these outliers in VMC and how to deal with them. This last point is the subject of *robust estimation*, a well-established field of statistics. Let us concentrate on the first point

### *Is the local energy distribution really Gaussian?*

Given a trial wave function $\Psi_T$ we can define the distribution of the local energy as

$$r(E) = \frac{\int \Psi_T(\mathbf{R}) d(E - E_L(\mathbf{R})) \Psi_T(\mathbf{R}) d\mathbf{R}}{\int \Psi_T^2(\mathbf{R}) d\mathbf{R}} \tag{13}$$

If the wave function is an exact eigenstate, the local energy is constant and the local energy distribution is a Dirac delta. If the trial wave function is sufficiently good, the local energy distribution is very sharp, located around the exact energy, and it can be well approximated by a Gaussian. However in the general case there really are no reasons why $r(E)$ should be well approximated by a Gaussian.

Let us consider two analytical examples first. Consider the trial wave function $\Psi_T = e^{-ar}$ for the Hydrogen atom. The local energy associated to this function is

$$E_L = -\frac{a^2}{2} + \frac{a-1}{r}. \tag{14}$$

Note that if $a > 1$  $E_L > -a^2/2$, while if $a < 1$  $E_L < -a^2/2$. This means that $r(E)$ is zero below or above $-a^2/2$. For $a = 1$ $E_L = -1/2$, the local energy distribution $r(E)$ is a Dirac delta and the trial wave function becomes the ground state eigenfunction.

Suppose that $a < 1$. Then equation (13) can be easily integrated, giving (not normalized)

$$r(E) = \begin{cases} \dfrac{(1-a)^3 a^3 e^{-\frac{4(a-1)a}{a^2+2E}}}{(a^2 + 2E)^4} & E < -a^2/2 \\ 0 & E \geq -a^2/2 \end{cases} \tag{15}$$

A similar formula can be obtained in the case $a > 1$. The plot for $a = 0.8$ is shown in Figure 1. First note that this function is neither a Gaussian, nor peaked on the exact eigenvalue of -1/2. The distribution is quite skewed, and it even contains an essential singularity. For values greater than $-a^2/2$, the distribution is zero. On the other hand, the left tail decays as $E^{-4}$: substantially slower than Gaussian decay. This means that the number of outliers observed will be higher than predicted by the Gaussian model.

Another interesting case is the 3D harmonic oscillator. Let us consider the trial function $\Psi_T = e^{-ar^2}$ where, for a = 1/2 the exact wave function is recovered. For the case $0 < a < \dfrac{1}{2}$ Eq. 13 can be integrated, giving

$$r(E) = \begin{cases} 0 & E \leq 3a \\ \dfrac{8a}{1-4a^2} \sqrt{\dfrac{4a(E-3a)}{p(1-4a^2)}} e^{-\frac{4a(E-3a)}{1-4a^2}} & E > 3a \end{cases} \qquad (16)$$

Even in this case the local energy distribution is far from being a Gaussian. The tail is exponentially decaying, so even in this case, in a Monte Carlo simulation, we should expect a number of "bad" walkers greater than what a simple Gaussian distribution predicts.

Only for very simple cases the analytical form of the local energy distribution can be obtained, so it is not possible to make a general statement on its form. However, it seems clear, even from numerical evidence [20,23,24] that the assumption of a Gaussian tail, implicit in the variance minimization, might not be a good choice, since in this case the "bad" walkers can spoil the entire optimization process, especially if the trial function is not flexible enough.

## *A robust optimization procedure*

A possible solution to this problem comes from the field of *Robust Estimation*. If the outlier points are troublesome, a sensible way to proceed could be to take into account their presence from the very beginning, assuming a different functional form for the tails of the local energy distribution. Instead of starting from a simple Gaussian distribution (Eq. 11) we can assume a distribution with higher tails. This approach is called the *M-estimates* approach.

For example, we could assume that the local energy distribution is exponentially decaying

$$P_i \propto e^{-|E_L(\mathbf{R}_i) - E_R|} \qquad (17)$$

Here the tails of the distribution, although exponentially decaying, are asymptotically much higher than those of a Gaussian. Using simple algebra it is easy to show that this assumption leads to the minimization of the sum of the absolute values

$$\min_{\mathbf{a}} \sum_{i}^{N} |E_L(\mathbf{R}_i, \mathbf{a}) - E_R| \qquad (18)$$

In this way a lower weight is given to the outliers and the risk of spoiling the optimization is reduced. In this way we have derived one of the functionals studied by Alexander *et al.*[22] and empirically found to cure the problem of false convergence by Lester and coworkers[31] in electronic structure calculations.

Of course it would be possible to start with a distribution with an even higher tail, like the Cauchy distribution, obtaining

$$\min_{\mathbf{a}} \sum_{i}^{N} \log\left[1 + (E_L(\mathbf{R}_i; \mathbf{a}) - E_R)^2 / 2\right] \qquad (19)$$

Even better would be to use the "real" local energy distribution for a given trial wave function model. For example, in the hydrogen atom example, the tails of the distribution should behave as $E^{-4}$. However, one rarely knows the functional form of the local energy distribution for the trial function employed, but it should be sufficient to use any distribution with sufficiently high tails to reduce or eliminate the problems caused by the outliers.

## A REAL EXAMPLE: THE HELIUM TRIMER

The optimization of good trial functions is an important issue in the study of pure and doped Helium clusters, and in the field of weakly bound clusters in general. The functional forms usually employed[24-28] are not very flexible and this frequently generates problems like those we have described[24,26,27]. A particularly problematic system is the Helium trimer, due to its very diffuse nature and high anharmonicity. The optimization of trial functions for the Helium trimer has been reported to cause problems[24,26], so it is a good testing ground for our investigation. The local energy distribution for a common trial wave function[29] for $^4He_3$, computed using VMC, is shown in Figure 2. We employed the LM2M2 potential[30], but the results do not depend on the particular form of the potential used. The local energy distribution has a noticeably non-Gaussian shape. Although most of the curve lies in the negative energy region, there is a sizable portion in the positive part. More problematic is the fact that there is a very slowly decaying tail for energies very far from the average. In this case, the optimization of the variance of the local energy is likely to cause problems and probably it would be better to assume a more slowly decaying distribution. To test our proposal, we optimized the sum of the absolute deviations and the sum of the square deviations.

We performed two sets of five optimization cycles, starting from the same trial wave function and from the same ensemble of 5000 walkers. In the first set we optimized the variance of the local energy, while in the second set we optimized the absolute deviation. After each optimization step the energy was computed by a VMC simulation. The newly generated ensemble was used as fixed sample for the next optimization.

The calculated energies are reported in Figure 3. As expected, the variance minimization is troublesome. The variational energy of the starting trial wave function is $-0.0725(1)$ cm$^{-1}$. Although the variance minimization procedure is able to produce, after four optimization cycles, a trial wave function slightly better than the starting one, giving $-0.0747(1)$ cm$^{-1}$, its trend is erratic, and even generates very bad trial wave functions during the process. On the other end, minimizing the absolute deviation seems to be a much more reliable procedure, showing a much smoother trend. Furthermore, the final wave function gives a much better energy of $-0.0805(1)$ cm$^{-1}$. We tried optimizing other objective functions, including Eq. 19, all sharing a less fast decaying tail than a Gaussian, obtaining similar results. This means that minimizing the absolute deviation is not the only possible choice here, and the good behavior is not directly related to some peculiar property of the absolute deviation function, but rather to the assumption of a less fast decaying tail. There might be cases where even an exponentially decaying tail might be too fast, and some benefit could be gained in assuming a power-like decaying tail.

## CONCLUSIONS

In this paper we have reexamined the discrete nature of the energy variance minimization algorithm in quantum Monte Carlo methods from the *maximum-likelihood* point of view, without regards to any integral approximation. This allowed us to unveil the origin of the problem of false convergence and to suggest alternative procedures designed to reduce or eliminate the problem. We tested our proposal optimizing a trial wave function for the Helium trimer using several algorithms. The minimization of the absolute deviation is shown to be more reliable than the variance minimization, void of convergence problems, and able to generate trial wave functions with better variational energy.

## ACKNOWLEDGMENTS

We would like to thank Peter Reynolds for discussions and carefully reading of the manuscript.

**FIGURE CAPTIONS**

Figure 1: Local energy distribution for the hydrogen atom with $\Psi_T = e^{-0.8r}$ (see text)

Figure 2: Local energy distribution for the helium trimer wave function (see text)

Figure 3: Sequence of optimization cycles for the helium trimer wave function

Figure 1
Bressanini et al.
Journal of Chemical Physics

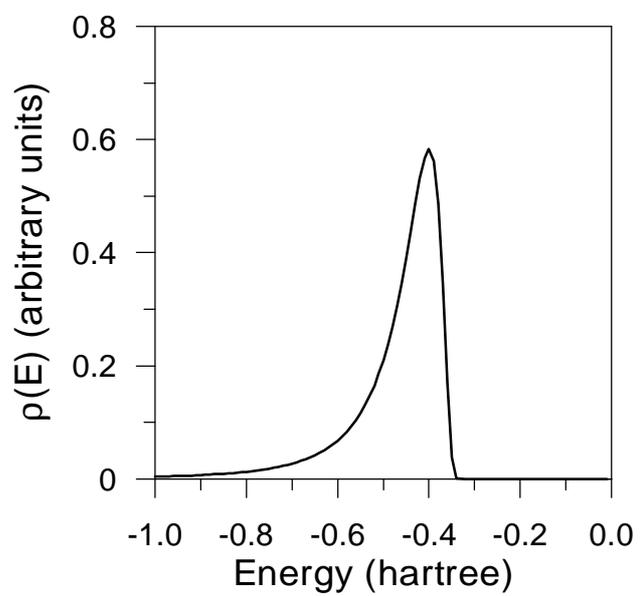

Figure 2
Bressanini et al.
Journal of Chemical Physics

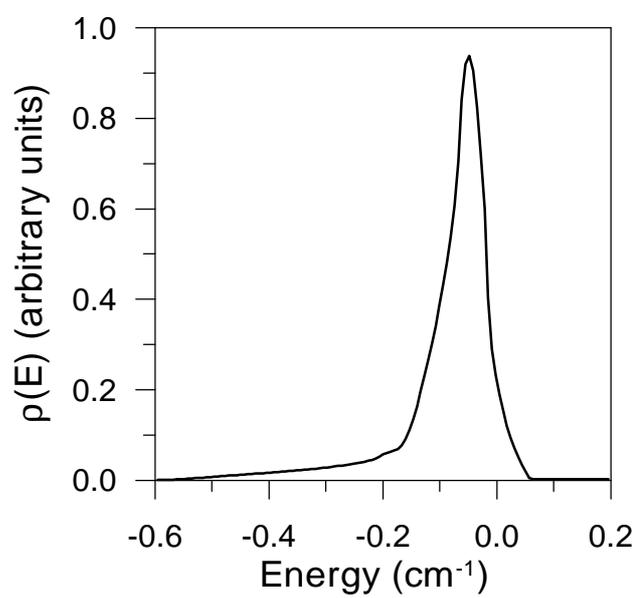

Figure 3
Bressanini et al.
Journal of Chemical Physics

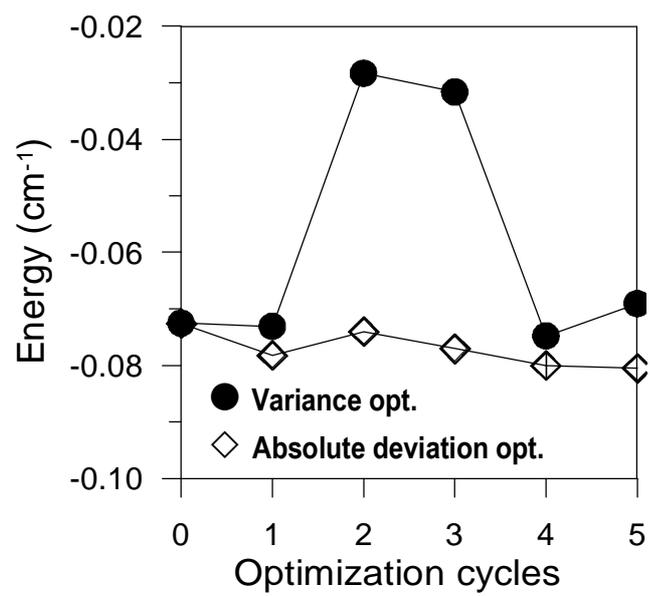